\documentclass[preprint2]{aastex62}

\usepackage{float}

\shorttitle{Six Years of Sustained Activity from Active Asteroid (6478)~Gault}
\shortauthors{Chandler et. al}

\usepackage{graphicx}

\begin{document}

\title{Six Years of Sustained Activity from Active Asteroid (6478)~Gault}

\correspondingauthor{Colin Orion Chandler}
\email{orion@nau.edu}

\author[0000-0001-7335-1715]{Colin Orion Chandler}
\affiliation{Department of Physics \& Astronomy, Northern Arizona University, PO Box 6010, Flagstaff, AZ 86011, USA}

\author[0000-0001-8531-038X]{Jay Kueny}
\affiliation{Department of Physics \& Astronomy, Northern Arizona University, PO Box 6010, Flagstaff, AZ 86011, USA}

\author[0000-0002-7600-4652]{Annika Gustafsson}
\affiliation{Department of Physics \& Astronomy, Northern Arizona University, PO Box 6010, Flagstaff, AZ 86011, USA}

\author[0000-0001-9859-0894]{Chadwick A. Trujillo}
\affiliation{Department of Physics \& Astronomy, Northern Arizona University, PO Box 6010, Flagstaff, AZ 86011, USA}

\author[0000-0002-3196-414X]{Tyler D. Robinson}
\affiliation{Department of Physics \& Astronomy, Northern Arizona University, PO Box 6010, Flagstaff, AZ 86011, USA}

\author[0000-0003-4580-3790]{David E. Trilling}
\affiliation{Department of Physics \& Astronomy, Northern Arizona University, PO Box 6010, Flagstaff, AZ 86011, USA}


\begin{abstract}
\label{Abstract}
We present archival observations demonstrating that main belt asteroid (6478)~Gault has an extensive history of comet-like activity. Outbursts have taken place during multiple epochs since 2013 and at distances extending as far as ~2.68 au, nearly aphelion. (6478)~Gault is a member of the predominately S-type (i.e., volatile-poor) Phocaea family; no other main belt object of this type has ever shown more than a single activity outburst. Furthermore, our data suggest this is the longest duration of activity caused by a body spinning near the rotational breakup barrier. If activity is indeed unrelated to volatiles, as appears to be the case, (6478) Gault represents a new class of object, perpetually active due to rotational spin-up.
\end{abstract}


\section{Introduction}
\label{introduction}

\begin{figure}[H]
	\centering
	\includegraphics[width=1.0\linewidth]{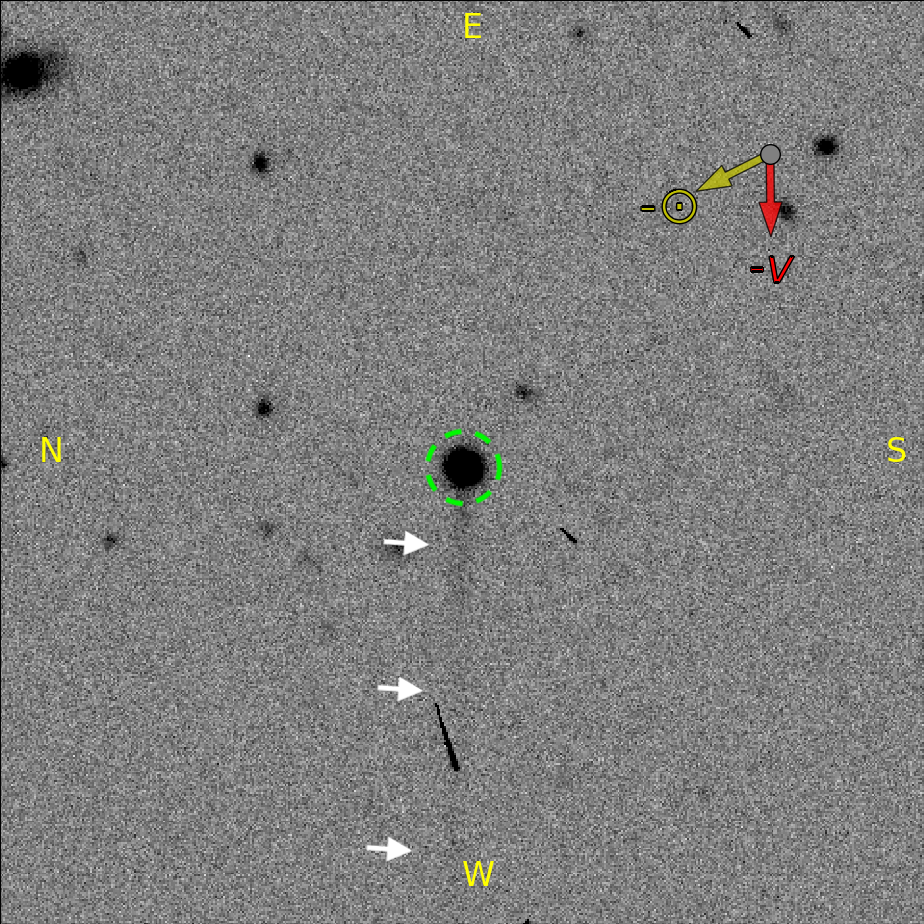}
    \caption{(6478)~Gault (dashed green circle) displays a prominent tail (indicated by white arrows) during this 2013 September 28  apparition when (6478)~Gault was halfway between perihelion and aphelion. This 90 s $g$-band exposure reached $\sim$7 mag fainter than (6478)~Gault. The anti-Solar direction ($-\odot$; yellow) and negative heliocentric velocity vector ($-\vec{v}$; red) are shown.}
    \label{fig:gault2013a}
\end{figure}

Active asteroids like (6478)~Gault (Figure~\ref{fig:gault2013a}, this work) are dynamically asteroidal objects but they uncharacteristically manifest cometary features such as tails or comae \citep{Hsieh:2006jt}. With only $\sim$20 known to date (see Table 1 of \citealt{Chandler:2018hw}), active asteroids remain poorly understood, yet they promise insight into solar system volatile disposition and, concomitantly, the origin of water on Earth \citep{Hsieh:2006dk}.

Active asteroids are often defined as objects with (1) comae, (2) semimajor axes interior to Jupiter, and (3) Tisserand parameters with respect to Jupiter $T_\mathrm{J}>3$; $T_\mathrm{J}$ describes an object's orbital relationship to Jupiter by 

\begin{equation}
    T_\mathrm{J} = \frac{a_\mathrm{J}}{a} + 2 \sqrt{\frac{a\left(1-e^2\right)}{a_\mathrm{J}}}\cos\left(i\right)
\end{equation}

\noindent where $a_\mathrm{J}$ is the semimajor axis of Jupiter, $e$ is the eccentricity, and $i$ is the inclination; see \cite{Jewitt:2012hv} for a thorough treatment. Main belt comets are a subset of active asteroids dynamically constrained to the main asteroid belt and thought to have volatile-driven activity (see e.g., \citealt{Snodgrass:2017kg} for an in-depth discussion). It is worth pointing out that some objects have had multiple classifications, for instance (3552) Don Quixote has an asteroid designation due to its low activity and has been called a near-Earth asteroid \citep{Mommert:2014gl} but has a $T_\mathrm{J}$ of 2.3 which indicates it is more properly a Jupiter family comet. 

Discovering these objects has proven observationally challenging. The first active asteroid, (4015)~Wilson-Harrington, was discovered in 1949 \citep{Harris:1950ub}. In the mid-1980s a connection between bow-shock magnetic field disturbances detected by the \textit{Pioneer} spacecraft suggested (2201)~Oljato was leaving behind a distant comet-like gas trail \citep{Kerr:1985gp}, even if not detected at the object itself \citep{Russell:1984eq}. Despite many efforts (see e.g., \citealt{Chamberlin:1996gx}) it was not until the 1996 discovery of activity in (7968)~Elst-Pizarro that another active asteroid was visually identified \citep{Elst:1996vj}. Though initially impact appeared a possible cause (e.g., \citealt{Toth:2000wm}), when activity recurred \citep{Hsieh:2010fm} it was more indicative of volatile sublimation.

A significant complication hindering our understanding of active asteroids arises when assessing underlying activity mechanisms: causes are neither few nor mutually exclusive (see \citealt{Jewitt:2012hv} for a comprehensive overview). Responsible primary processes include volatile sublimation (e.g., 133P/Elst-Pizarro,  \citealt{Hsieh:2010fm}), impact events (e.g., (596) Scheila,  \citealt{Bodewits:2011fo,Moreno:2011ki}), rotational breakup (e.g., 311P/PanSTARRS, \citealt{Jewitt:2013id,Moreno:2014gf}), thermal fracture (e.g., (3200) Phaethon, discussed below), and cryovolcanism (e.g., (1) Ceres, \citealt{Kuppers:2014dp,Witze:2015hq}). Physical interaction, or ``rubbing binary,'' has been proposed as a primary mechanism in the case of 311P (\citealt{Hainaut:2014im}, cf. \citealt{Jewitt:2018gm}). Secondary mechanisms such as electrostatic gardening, physical properties like chemical makeup, and geometric effects (e.g., the opposition effect) may influence our ability to reliably detect and quantify outbursts.

One crucial diagnostic indicator of the underlying activity mechanism is whether or not activity recurs. If activity is observed on only one occasion (i.e., a single apparition), then the object may have experienced a recent impact event. Expelled material and/or exposed volatiles sublimating may both cause comae or tails to appear. Activity would then cease once ejecta dissipated or the volatile supply is exhausted, reburied, or refrozen.

Recurrent activity is typically associated with volatile sublimation. For example, Geminid Meteor Shower parent body (3200) Phaethon is thought to undergo thermal fracture during the rapid temperature changes accompanying its perihelion passages \citep{Li:2013fs} where it experiences temperatures $>800$ K \citep{Ohtsuka:2009ky}. Fracture events may directly expel material in addition to exposing volatiles for sublimation.

Thermally induced activity is thought to increase with decreasing heliocentric distance; that is, the closer a body is to the Sun, the more likely an outburst is to occur. 
Active asteroids are more likely to exhibit activity during perihelion passage (see Table 1 of \citealt{Chandler:2018hw}). Notable exceptions where activity was discovered at distances far from perihelion include 311P/PanSTARRS \citep{Jewitt:2013id} and (493) Griseldis \citep{Tholen:2015tu}. Activity in ``traditional'' comets has been reported at distances that are substantially farther than the main asteroid belt, for instance Comet C/2010 U3 (Boattini) at 27~au \citep{Hui:2019ey}. Of the $\sim$20 active asteroids known to date, 16 are carbonaceous (i.e., C-type) but only four are believed to be composed of silicate-rich non-primitive material (i.e., \citealt{DeMeo:2009gz} S-type taxonomy): (2201) Oljato (Apollo-orbit), 233P/La Sagra (Encke-orbit), 311P/PanSTARRS (inner main belt), and 354P/LINEAR (outer main belt). 

(6478)~Gault activity was first reported in 2019 January \citep{Smith:2019wy}. Ensuing follow-up observations \citep{Maury:2019wy} confirmed activity with subsequent reports \citep{Lee:2019wi,Ye:2019tr} providing evidence of ongoing activity. \cite{Ye:2019ck} reported that multiple outbursts actually began in 2018 December. Analysis of dust emanating from (6478)~Gault via Monte Carlo tail brightness simulations indicate the current apparition, comprised of two outbursts, could have begun as early as 2018 November 5 \citep{Moreno:2019vp}. Simultaneous to our own work, \cite{Jewitt:2019vg} reported three tails with independent onsets, the earliest being 2018 October 28.

We set out to determine if any data in our local repository of National Optical Astronomy Observatory (NOAO) Dark Energry Camera (DECam) images showed signs of activity. The $\sim$500 megapixel DECam instrument on the Blanco 4 m telescope situated on Cerro Tololo, Chile, probes faintly ($\sim$24 mag) and, as we demonstrated in \cite{Chandler:2018hw}, it is well-suited to detect active asteroids. We produced novel tools taking into account (1) orbital properties of (6478)~Gault (summarized in Appendix \ref{sec:GaultData}) and (2) observational properties (e.g., apparent magnitude, filter selection, exposure time) to find ideally suited candidate images.


\section{Methods}
\label{sec:methods}

We searched our own in-house database of archival astronomical data (e.g., observation date, coordinates) in order to locate images that are likely to show (6478)~Gault. Our database includes the entire NOAO DECam public archive data tables along with corresponding data from myriad sources (e.g., NASA JPL Horizons \citealt{Giorgini:1996ts}; see also the Acknowledgements).

\subsection{Locating Candidate Images}
\label{subsec:candidates}

We began our search for (6478)~Gault by making use of a fast grid query in R.A. and decl. space. We then passed these results through a more accurate circular filter prescribed for the DECam image sensor arrangement. Lastly, we computed image sensor chip boundaries precisely to ensure that the object fell on a sensor rather than, for example, gaps between camera chips. This progressively more precise query approach cut down image search time by orders of magnitude.

\subsection{Observability Assessment}
\label{subsec:observability}

We created a reverse exposure time calculator to estimate how faintly (i.e., the magnitude limit) candidate images probed. After applying color coefficient corrections (see \citealt{Willmer:2018hd} for procedure details) we transformed the color-corrected magnitudes to the absolute bolometric system used by the DECam exposure time calculator\footnote{\url{http://www.ctio.noao.edu/noao/node/5826}}. These steps enabled us to compute differences between apparent magnitude and the specific magnitude limit of the DECam exposure so that we could produce a list of images where (6478)~Gault could be detected.

\subsection{Thumbnail Extraction}
\label{subsec:acquisition}

We downloaded the image files containing (6478)~Gault from the NOAO archive and, following the procedures of \citet{Chandler:2018hw}, we extracted flexible image transport system (FITS) thumbnails of (6478)~Gault. We then performed image processing to enhance contrast before finally producing portable network graphics (PNG) image files for inspection.

\subsection{Image Analysis}
\label{subsec:imageanalysis}

We visually inspected our (6478)~Gault thumbnail images to check for signs of activity. PNG thumbnails with activity indicators were examined in greater detail via the corresponding FITS thumbnail image.

To assess the influence of heliocentric distance on activity level we employed a simple metric (see \cite{Chandler:2018hw} ``$\mathrm{\%}_\mathrm{peri}$'' for motivation) describing how far from perihelion $q$ the target $T$ was located (at distance $d$) relative to its aphelion distance $Q$ by

\begin{equation}
		\%_{T\rightarrow q} = \left(\frac{Q - d}{Q-q}\right)\cdot 100\mathrm{\%}.
		\label{eq:percentperi}
\end{equation}

\vspace{2mm}


\section{Results}
\label{sec:results}

\begin{table*}
\caption{(6478)~Gault Archival DECam Observations Examined}
\footnotesize
\centering
\begin{tabular}{ccrlrcccccrr}
Activity & UT Date & Time & Processing  & $t_\mathrm{exp}$ & Filter & $m_\mathrm{lim}$ & $m_V$ & $\Delta m$ & $r$ & $\%_{T\rightarrow q}$ & $\angle_\mathrm{STO}$  \\
&                        &                       &           &           (s)            &           &           &                   &   &   (au) &            &   ($^\circ$)\\
\hline\hline
 & 2013 Sep 22               & 3:03                     & R, I, Re    & 45                               & Y          & 21.0                                 & 17.2                     & -4                           & 2.27                          & 54                        & 3.41                            \\
$*$ & 2013 Sep 28               & 2:23                     & R, I, I, Re & 90                               & g          & 24.2                               & 17.0                       & -7                           & 2.28                          & 52                        & 0.45                            \\
$*$ & 2013 Oct 13               & 3:06                     & I, I, I, Re & 90                               & i          & 23.3                               & 17.1                     & -6                           & 2.32                          & 49                        & 8.21                            \\
& 2013 Oct 13               & 3:08                     & I, I, Re    & 90                               & z          & 22.7                               & 17.1                     & -6                           & 2.32                          & 49                        & 8.21                            \\
$*$ & 2016 Jun 09               & 4:45                     & R, I, Re    & 96                               & r          & 23.9                               & 16.8                     & -7                           & 1.86                          & 100                           & 22.38                           \\
$*$ & 2016 Jun 10               & 4:40                     & R, I, Re    & 107                              & g          & 24.2                               & 16.8                     & -7                           & 1.86                          & 100                           & 22.48                           \\
 & 2017 Oct 23               & 8:57                     & R, I, Re    & 80                               & z          & 22.7                               & 18.8                     & -4                           & 2.66                          & 10                         & 14.79                           \\
 & 2017 Nov 11               & 7:13                     & R, I, Re    & 80                               & z          & 22.7                               & 18.5                     & -5                           & 2.68                          & 8                        & 10.92                           \\
$*$ & 2017 Nov 12               & 5:14                     & R, I, Re    & 111                              & g          & 24.3                               & 18.5                     & -6                           & 2.68                          & 8                        & 10.81                          \\
\hline
\end{tabular}

\raggedright
\footnotesize
\vspace{1mm}
\textbf{Note.} Process types: Raw (R), InstCal (I), Resampled (Re); $r$: Sun-target distance; $\%_{T\rightarrow q}$: target distance toward perihelion from aphelion (Equation \ref{eq:percentperi}); $t_\mathrm{exp}$: exposure time; $m_\mathrm{lim}$: estimated exposure magnitude limit; $m_V$: (6478)~Gault apparent $V$-band magnitude; $\Delta m$: $m - m_\mathrm{lim}$; $\angle_\mathrm{STO}$: Sun-Target-Observer (phase) angle. 
 Thumbnails are included in Appendix \ref{sec:ThumbnailGallery}.
\label{tab:observations}
\end{table*}

\begin{figure*}
    \centering
    \begin{tabular}{cc}
    \includegraphics[width=0.5\linewidth]{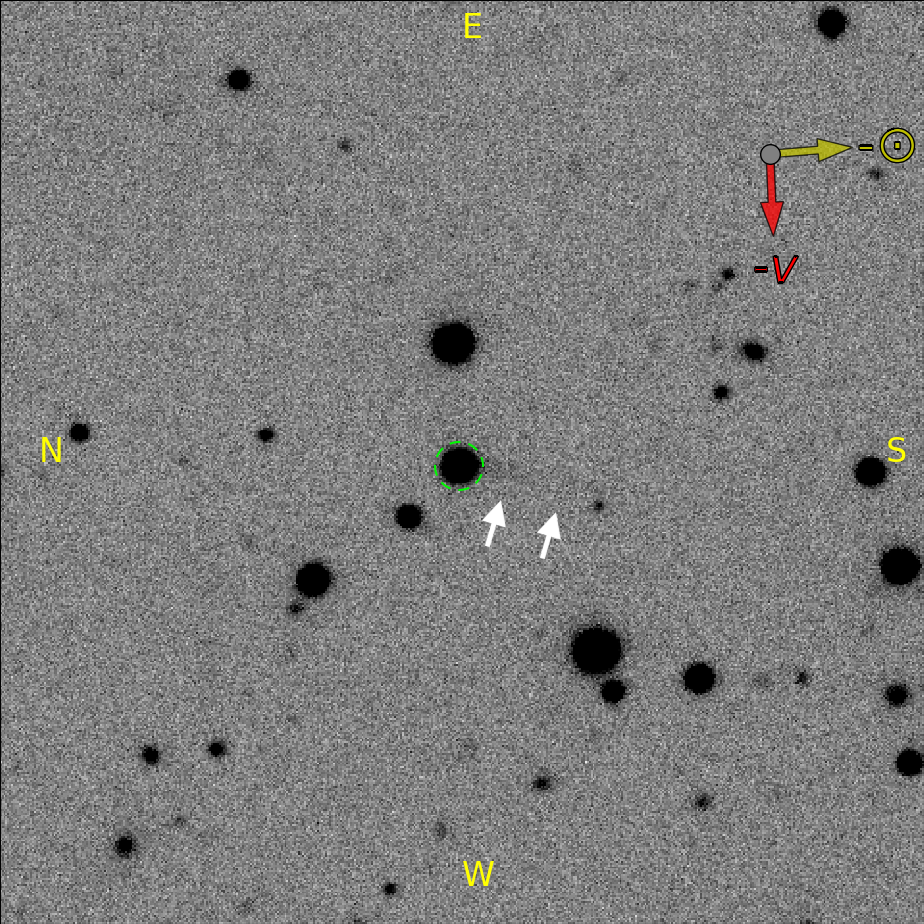}  &  \includegraphics[width=0.5\linewidth]{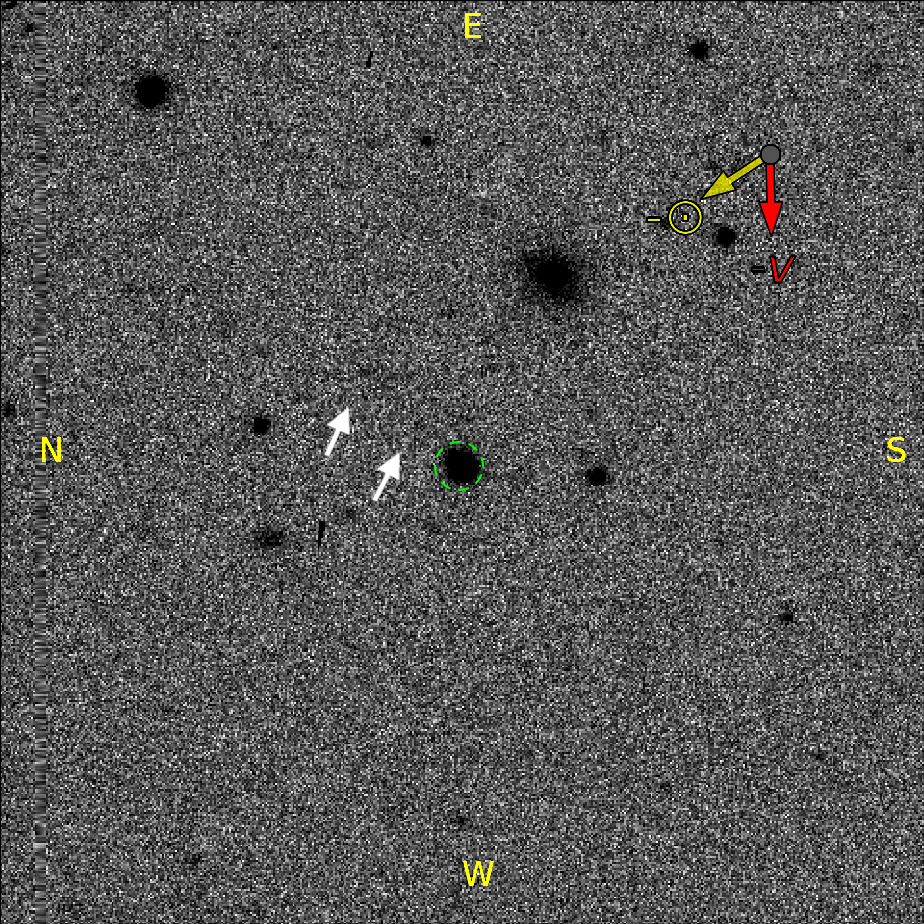}\\
    \textbf{(a)}     & \textbf{(b)}\\
    \end{tabular}
    \caption{\textbf{(a)} A tail (indicated by white arrows) at $\sim$8 o'clock is seen in this 107-second $g$-band exposure imaged June 10, 2016. The yellow arrow indicates anti-Solar ($-\odot$) direction and red the negative heliocentric velocity vector ($-\vec{v}$). \textbf{(b)} (6478)~Gault top seen on November 12, 2017. The 111 second exposure in the $g$-band delivered a flux limit 6 magnitudes fainter than (6478)~Gault, revealing a faint tail ($\sim$2 o'clock, indicated by white arrows) and coma. The yellow arrow indicates anti-Solar ($-\odot$) direction and the red arrow negative heliocentric velocity ($-\vec{v}$). Dashed green circles outline (6478)~Gault and white arrows have been placed perpendicular to any observed activity.}
    \label{fig:thumb2016}
\end{figure*}

We successfully extracted thumbnails from 9 archival observations of (6478)~Gault; see Table \ref{tab:observations} for details. Most data were available in raw and calibrated form (``InstCal'' and ``Resampled'' are described by \citealt{DarkEnergySurveyCollaboration:2014uh}) allowing us to extract $\sim$30 thumbnail images in total. Figure~\ref{fig:gault2013a} shows (6478)~Gault in 2013 with a pronounced tail in the 6 o'clock direction. Figure~\ref{fig:thumb2016} \textbf{(a)} and Figure~\ref{fig:thumb2016} \textbf{(b)} show (6478)~Gault in 2016 and 2017, respectively. Additional images may be found in Appendix \ref{sec:ThumbnailGallery}.

\begin{figure*}[ht]
	\centering
	\includegraphics[width=1.0\linewidth]{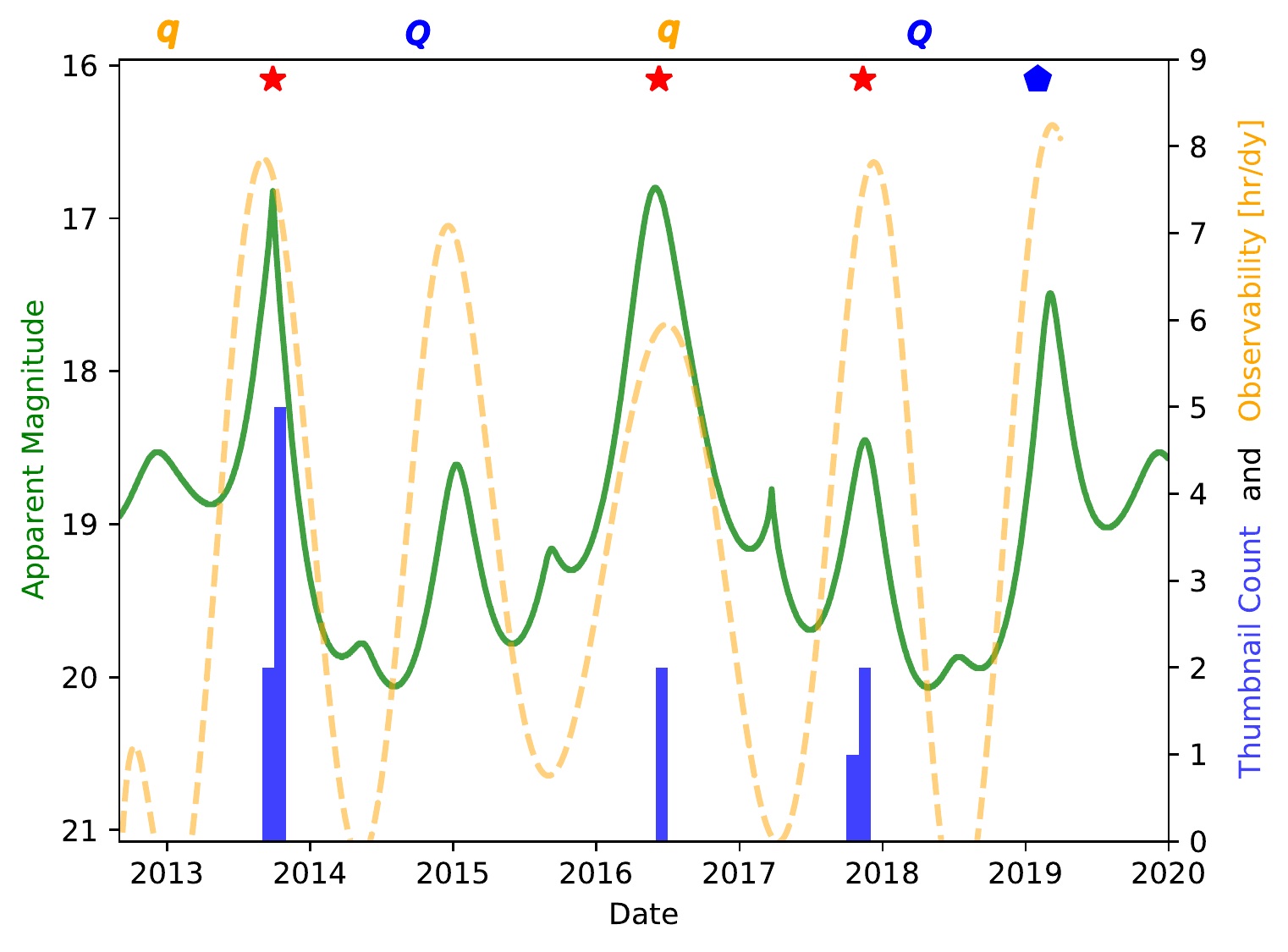}
	\caption{(6478)~Gault activity timeline beginning with DECam operation commencement (2012 September) to present. Red stars show when we found visible activity; the blue pentagon represents the current apparition where prominent activity has been seen. Above the top axis are marked perihelion ($q$) and aphelion ($Q$) events. The solid green line indicates the apparent $V$-band magnitude of (6478)~Gault as viewed from Earth. The dashed yellow line shows our ``observability'' metric, defined as the number of hours per UT observing date meeting both of the following conditions possible for DECam: (1) elevation $> 15^\circ$, and (2) the refracted solar upper-limb elevation was $< 0^\circ$ (i.e., nighttime). Peaks in apparent magnitude coinciding with peaks in observability indicate opposition events; conversely, secondary magnitude peaks aligned with observability troughs highlight solar conjunctions, i.e., when (6478)~Gault was ``behind'' the Sun as viewed from Earth. All activity has been observed near opposition events. Also, activity was seen at every epoch in our data. The histogram (vertical blue bars) indicate the number of thumbnails that we extracted for a given observing month.}
	\label{fig:ActivityTimeline}
\end{figure*}

\begin{figure}
    \centering
    \includegraphics[width=1.0\linewidth]{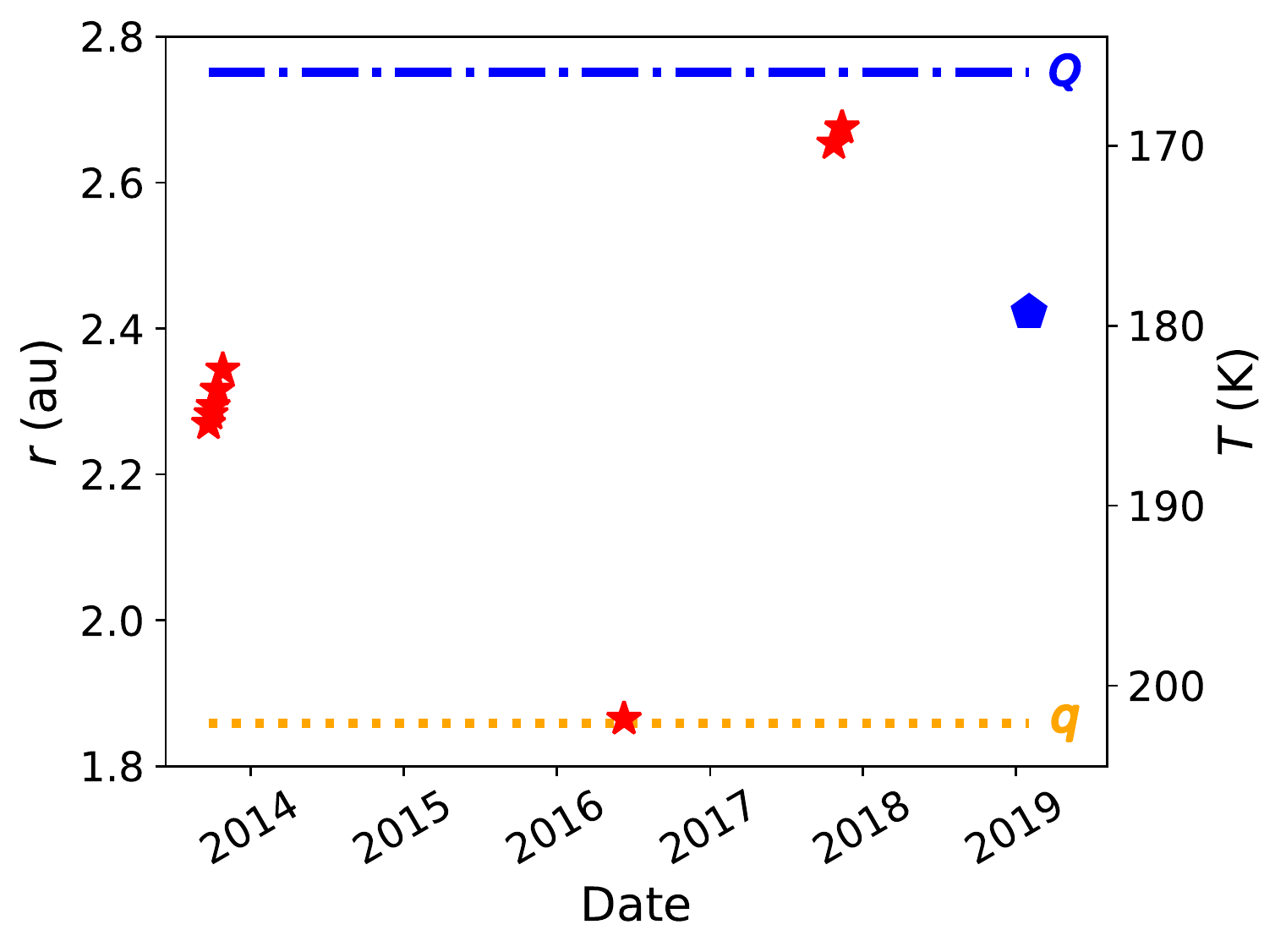}
    \caption{Positive detections of (6478)~Gault activity with DECam as a function of heliocentric distance $r$ (au) and surface temperature $T$ (K). Our activity observations are indicated by red stars, whereas the current apparition is represented by the blue pentagon. Distance and temperature of (6478)~Gault  perihelion $q$ (orange dashed line) and aphelion $Q$ (blue dashed-dotted line) events are shown. During the course of one full orbit, (6478)~Gault is exposed to temperatures greater than 165~K. As a result, (6478)~Gault is consistently subjected to temperatures that are too high for water ice to form at the 5 au ice formation distance \citep{Snodgrass:2017kg}.}
    \label{fig:tempsanddistance}
\end{figure}

Figure~\ref{fig:ActivityTimeline} summarizes our observed activity. We found activity at least once in every set of observations and no correlation with distance. We plotted apparent $V$-band magnitude (solid green line) and found that all periods of activity were observed near opposition events. We further define ``observability'' (dashed yellow line) as when the object was (1) above 15$^\circ$ elevation, and (2) visible outside of daylight hours. This allowed us to assess potential observational biases specific to the southern hemisphere where our data were collected. As demonstrated by the coinciding of apparent magnitude maxima with spikes in apparent magnitude, the primary observability factor was solar elongation.

Figure~\ref{fig:tempsanddistance} shows how (6478)~Gault varies in both temperature $T$ and distance $r$ through time. Indicated are our activity observations (red stars) and the current apparition (blue pentagon). Temperature varies between $\sim$165 K at aphelion $Q$ (blue dash-dotted line) and $\sim$200K at perihelion $q$ (orange dotted line).

We define persistent activity as activity detectable across contiguous sets of observations spanning at least two epochs, even if activity is not visible in every image (due to, for example, exposure time and/or filter selection). We also expect to see activity at all positions throughout the orbit where (6478)~Gault may be detectable by DECam, given appropriate observing parameters (e.g., exposure time, filter selection).


\section{Discussion} 
\label{sec:discussion}

Most active asteroids, like comets, are composed of low-albedo (i.e.,~dark), primitive material allowing for sublimation or release of volatiles to occur when the body is heated during close passages with the Sun \citep{Hsieh:2018gm}. 

Of the $\sim$20 known active asteroids, four belong to the S-type asteroid taxonomy defining non-primitive silicate-rich material \citep{DeMeo:2009gz}. For these four objects, the causes of activity, when identifiable, are thought to be rotational breakup or impact. Rotational breakup and impact events are consistent with single apparitions or short-lived activity. Furthermore, \cite{Hsieh:2018kk} found that processed material bodies, such as S-types, are more likely to become active due to disruption, while primitive material bodies, such as C-types, can become active via multiple mechanisms due to their volatile abundances. 

(6478)~Gault has been identified as a core member of the Phocaea Family \citep{Knezevic:2003hw}. The Phocaea family is dominated by 75\% S-types, followed by 15\% C-types, and 10\% a mix of other asteroid taxonomies \citep{Carvano:2001kb}. While this work was in review, \cite{Jewitt:2019vg} reported color measurements suggesting that (6478)~Gault is closer in taxonomic class to a C-type body, rather than an S-type. However, gases were not detected in their spectra, suggesting that sublimation may not be the underlying cause.


Sustained activity near perihelion normally can point to sublimation driven activity, but we observe activity nearly at aphelion during opposition. We do see variability in activity intensity, but we observe activity in (6478)~Gault in at least one image in each set of observations in our DECam data set, suggesting that the target is perpetually active. As a result, we conclude there is no correlation between distance and activity for (6478)~Gault.

Because we observe persistent activity, impact-driven disruption seems unlikely as we would expect the timescale for the activity to be relatively short. The most probable cause for activity has been presented as disruption due to rotational breakup of (6478)~Gault \citep{Moreno:2019vp, Ye:2019tr, Ye:2019ck} due to the Yarkovsky--O'Keefe--Radzievskii--Paddack (YORP) effect spin-up \citep{Kleyna:2019ip}; see \cite{McNeill:2016jf}, \cite{Lowry:2007by}, and \cite{Bottke:2006en} for detailed explanations of YORP forces.

Rotational breakup holds for an S-type composition where we would anticipate landslides or surface material redistribution caused by rapid rotation near the 2.2 hr spin rate barrier, which is consistent with the measured $\sim$2 hr light curve period reported by \cite{Kleyna:2019ip}. We predict that (6478)~Gault will continue to show signs of activity as it has for the last 6 years in a relatively steady state. 
We do not expect catastrophic disruption of (6478) Gault (cf.\ \citet{Moreno:2019vp}). 

The activity observed in (6478)~Gault over multiple epochs and throughout its orbit make (6478)~Gault the first known sustained-activity active asteroid in the main asteroid belt. As a likely S-type asteroid, this is also the first time that we have observed a sustained active body at the rotational barrier for such an extended duration. If activity is in fact not volatile-related, then Gault is a new class of object, perpetually active due to spin-up. We encourage continued monitoring of both the lightcurve and activity level of (6478) Gault, as well as photometric color observations or spectra to further explore its composition.


\section{Acknowledgements}
\label{sec:acknowledgements}

The authors thank the anonymous referee whose comments greatly improved the quality of this Letter.

We wish to thank Nick Moskovitz (Lowell Observatory), who strongly encouraged us to pursue this report. We thank Dr.\ Mark Jesus Mendoza Magbanua (University of California San Francisco) for his frequent and timely feedback on the project. Many thanks to David Jewitt and Man-To Hui (Univeristy of California Los Angeles) for their helpful comments. The authors express their gratitude to Jonathan Fortney (University of California Santa Cruz), Mike Gowanlock (NAU), Cristina Thomas (NAU), and the Trilling Research Group (NAU), all of whom provided invaluable insights which substantially enhanced this work. The unparalleled support provided by Monsoon cluster administrator Christopher Coffey (NAU) and his High Performance Computing Support team facilitated the scientific process wherever possible.

This material is based upon work supported by the National Science Foundation Graduate Research Fellowship Program under Grant No.\ (2018258765). Any opinions, findings, and conclusions or recommendations expressed in this material are those of the author(s) and do not necessarily reflect the views of the National Science Foundation.

Computational analyses were run on Northern Arizona University's Monsoon computing cluster, funded by Arizona's Technology and Research Initiative Fund. This work was made possible in part through the State of Arizona Technology and Research Initiative Program. ``GNU's Not Unix!'' (GNU) Astro \textit{astfits} \citep{Akhlaghi:2015gv} provided command-line FITS file header access.



This research has made use of data and/or services provided by the International Astronomical Union's Minor Planet Center. This research has made use of NASA's Astrophysics Data System. This research has made use of the The Institut de M\'ecanique C\'eleste et de Calcul des \'Eph\'em\'erides (IMCCE) SkyBoT Virtual Observatory tool \citep{Berthier:2006tn}. This work made use of the {FTOOLS} software package hosted by the NASA Goddard Flight Center High Energy Astrophysics Science Archive Research Center. This research has made use of SAO ImageDS9, developed by Smithsonian Astrophysical Observatory \citep{Joye:2003ua}. This work made use of the Lowell Observatory Asteroid Orbit Database \textit{astorbDB} \citep{Bowell:1994hm,Moskovitz:2018wi}. This work made use of the \textit{astropy} software package \citep{AstropyCollaboration:2013cd}.

This project used data obtained with the Dark Energy Camera (DECam), which was constructed by the Dark Energy Survey (DES) collaboration. Funding for the DES Projects has been provided by the U.S. Department of Energy, the U.S. National Science Foundation, the Ministry of Science and Education of Spain, the Science and Technology Facilities Council of the United Kingdom, the Higher Education Funding Council for England, the National Center for Supercomputing Applications at the University of Illinois at Urbana-Champaign, the Kavli Institute of Cosmological Physics at the University of Chicago, Center for Cosmology and Astro-Particle Physics at the Ohio State University, the Mitchell Institute for Fundamental Physics and Astronomy at Texas A\&M University, Financiadora de Estudos e Projetos, Funda\c{c}\~{a}o Carlos Chagas Filho de Amparo, Financiadora de Estudos e Projetos, Funda\c{c}\~ao Carlos Chagas Filho de Amparo \`{a} Pesquisa do Estado do Rio de Janeiro, Conselho Nacional de Desenvolvimento Cient\'{i}fico e Tecnol\'{o}gico and the Minist\'{e}rio da Ci\^{e}ncia, Tecnologia e Inova\c{c}\~{a}o, the Deutsche Forschungsgemeinschaft and the Collaborating Institutions in the Dark Energy Survey. The Collaborating Institutions are Argonne National Laboratory, the University of California at Santa Cruz, the University of Cambridge, Centro de Investigaciones En\'{e}rgeticas, Medioambientales y Tecnol\'{o}gicas–Madrid, the University of Chicago, University College London, the DES-Brazil Consortium, the University of Edinburgh, the Eidgen\"ossische Technische Hochschule (ETH) Z\"urich, Fermi National Accelerator Laboratory, the University of Illinois at Urbana-Champaign, the Institut de Ci\`{e}ncies de l'Espai (IEEC/CSIC), the Institut de Física d'Altes Energies, Lawrence Berkeley National Laboratory, the Ludwig-Maximilians Universit\"{a}t M\"{u}nchen and the associated Excellence Cluster Universe, the University of Michigan, the National Optical Astronomy Observatory, the University of Nottingham, the Ohio State University, the University of Pennsylvania, the University of Portsmouth, SLAC National Accelerator Laboratory, Stanford University, the University of Sussex, and Texas A\&M University.

Based on observations at Cerro Tololo Inter-American Observatory, National Optical Astronomy Observatory (NOAO Prop. IDs 2012B-0001, PI: Frieman; 2014B-0404, PI: Schlegel), which is operated by the Association of Universities for Research in Astronomy (AURA) under a cooperative agreement with the National Science Foundation.

\appendix

\section{Gault Data}
\label{sec:GaultData}

For reference we provide essential information regarding (6478) below.

\begin{center}
	\textbf{Properties of (6478)~Gault}
	\begin{tabular}{lll}
		Parameter & Value & Source\\
		\hline\hline
		Discovery Date & 1988 May 12 & \citet{Schmadel:2003wz}\\
		Discovery Observers & C. S. \& E. M. Shoemaker & \citet{Schmadel:2003wz}\\
		Discovery Observatory & Palomar & \citet{Schmadel:2003wz}\\
		Activity Discovery Date & & \citet{Smith:2019wy}\\
		Alternate Designations & 1988 JC1, 1995 KC1 & NASA JPL Horizons\\
		Orbit Type & Inner Main Belt & NASA JPL Horizons\\
		Family & Phocaea (Core Member) & {\citet{Knezevic:2003hw}}\\
		Taxonomic Class & S & via Phocaea association\\
		Diameter & $D=4.5$ km & {\citet{Harris:2002ux}}\\ 
		Absolute $V$-band Magnitude & $H=14.4$ & NASA JPL Horizons\\
		Slope Parameter & $G=0.15$ & NASA JPL Horizons\\
		Orbital Period & $P=3.5$ yr & NASA JPL Horizons \\
		Semimajor Axis & $a=2.305$ au & NASA JPL Horizons\\
		Eccentricity & $e=0.1936$ & NASA JPL Horizons\\
		Inclination & $i=22.8113^\circ$ & NASA JPL Horizons\\
		Longitude of Ascending Node & $\Omega=183.558$ & Minor Planet Center\\
		Mean Anomaly & $M=289.349^\circ$ & Minor Planet Center\\
		Argument of Perihelion & $\omega=83.2676^\circ$ & NASA JPL Horizons\\
		Perihelion Distance & $q=1.86$ au & NASA JPL Horizons\\
		Aphelion Distance & $Q=2.75$ au & NASA JPL Horizons\\
		Tisserand Parameter w.r.t. Jupiter & $T_J=3.461$ & astorbDB\\
	\end{tabular}
\end{center}

\section{Thumbnail Gallery}
\label{sec:ThumbnailGallery}

For all thumbnails, red arrows indicate the negative motion vector $-\vec{v}$ of (6478)~Gault; yellow arrows point away from the Sun ($-\odot$). When possible, (6478)~Gault has been circled with a dashed green line and white arrows placed perpendicular to any observed activity. Areas outsize of chip boundaries appear black in color.

\begin{center}
\begin{tabular}{cc}
     \includegraphics[width=0.5\linewidth]{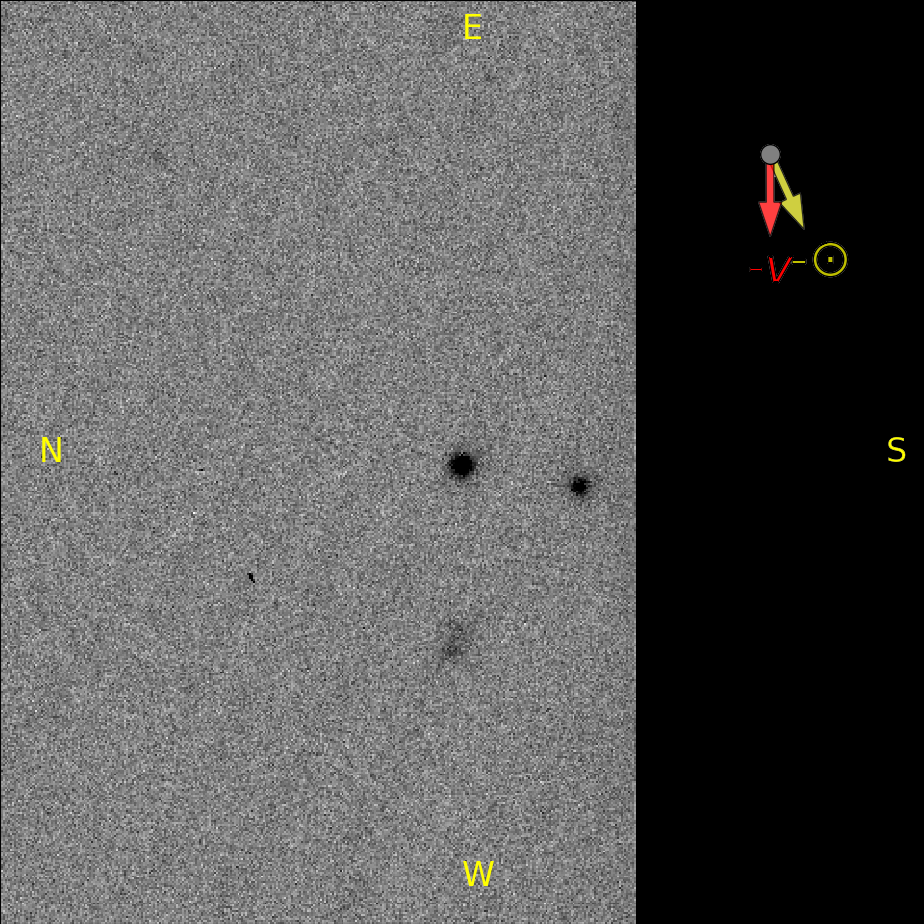} &  \includegraphics[width=0.5\linewidth]{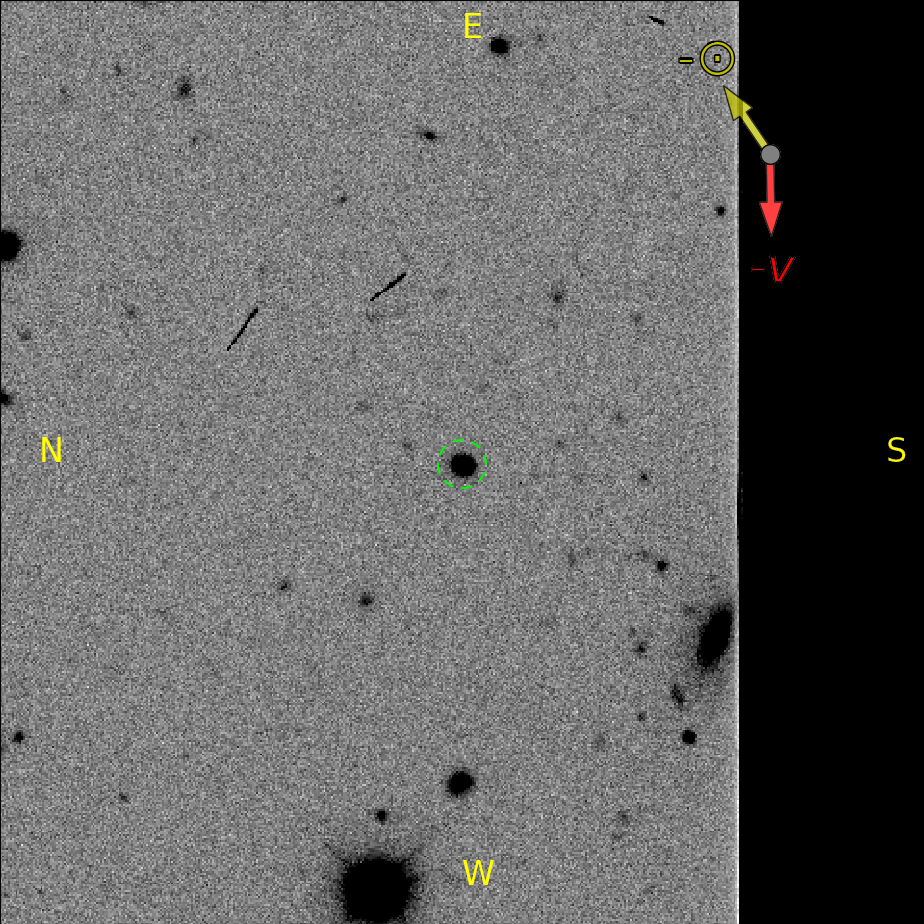}\\
\end{tabular}
\footnotesize
    Left panel: 2013 September 22 3:03 (UT); 45 s $Y$-band. Right panel: 2013 October 13 03:06 (UT); 90 s $i$-band.
\end{center}


\begin{center}
\begin{tabular}{cc}
    \includegraphics[width=0.5\linewidth]{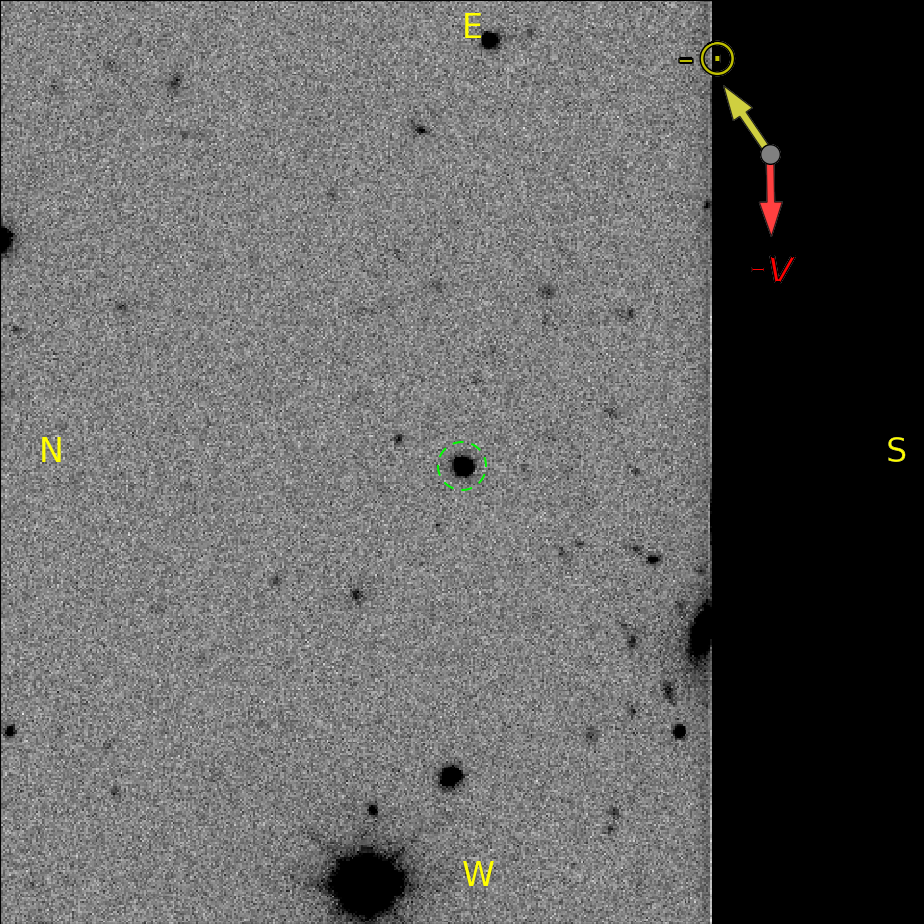} &
    \includegraphics[width=0.5\linewidth]{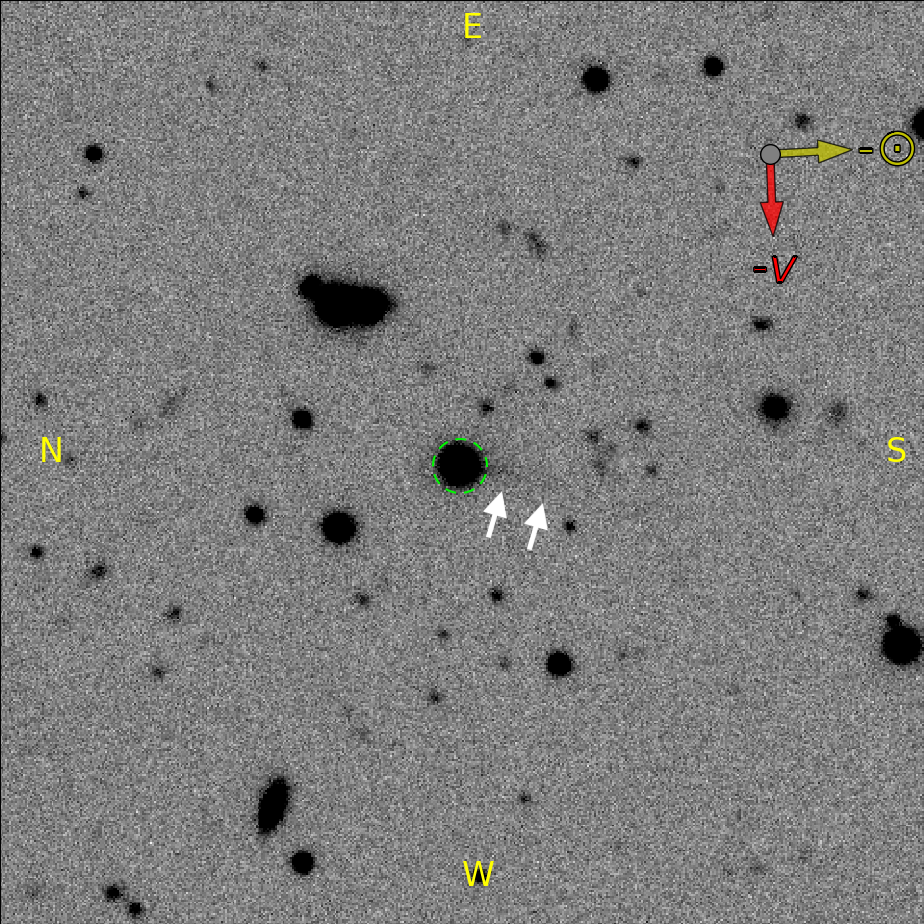} \\
\end{tabular}
\footnotesize
    Left panel: 2013 October 13 03:08 (UT); 90 s $z$-band. Right panel: 2016 June 09 04:45 (UT); 96 s $r$-band.
\end{center}

\begin{tabular}{cc}
     \includegraphics[width=0.5\linewidth]{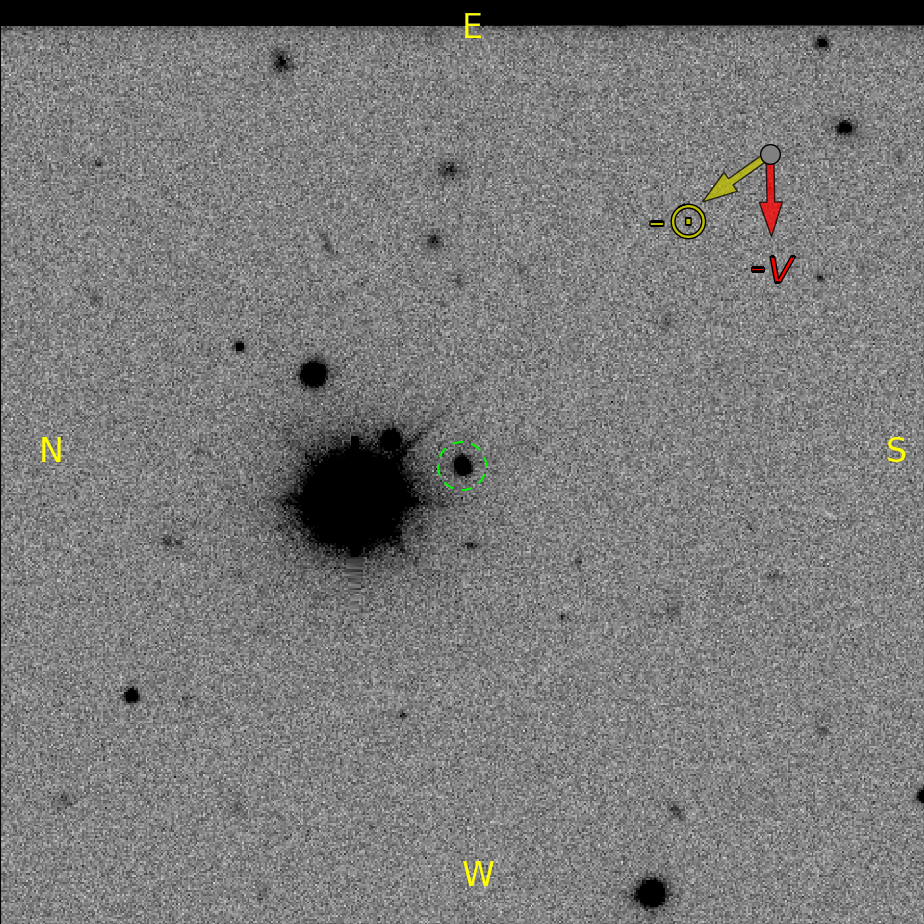} & \includegraphics[width=0.5\linewidth]{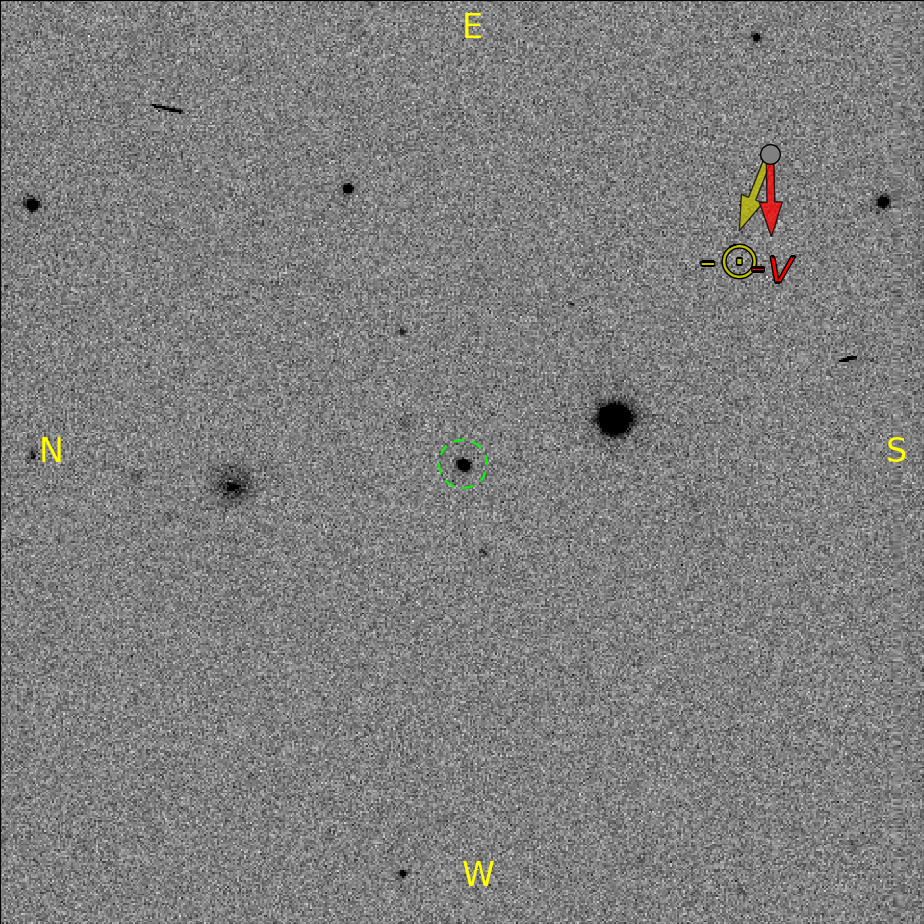} \\
\end{tabular}
\begin{center}
\footnotesize
    Left panel:  2017 November 11 07:13 (UT); 80 s $z$-band. Right panel: 2017 October 23 08:57 (UT); 80 s $z$-band.
\end{center}


\end{document}